\documentclass[amsmath,amssymb,superscriptaddress,nobalancelastpage,prb,twocolumn]{revtex4-1}
\usepackage{hyphenat}
\usepackage{graphicx,xcolor}
\usepackage{graphicx}
\usepackage{varioref}
\usepackage{xr-hyper}
\usepackage{xcolor}
\usepackage{nicefrac}
\usepackage{xfrac}
\usepackage{hyperref}
\hypersetup{colorlinks,linkcolor=blue,urlcolor=blue,citecolor=blue}
\usepackage{ulem}
\usepackage{siunitx}
\usepackage{graphicx}
\usepackage{dcolumn}
\usepackage{bm}
\usepackage{braket}
\usepackage{wasysym}
\usepackage{textcomp}
\usepackage{mhchem}

\newcommand{\cref}[1]{Chap.~\ref{#1}}
\newcommand{\tref}[1]{Tab.~\ref{#1}}
\newcommand{\fref}[1]{Fig.~\ref{#1}}

\newcommand{\lco}{La$_2$CuO$_4$}

\makeatletter
\newlength{\Laped}
\newcommand{\lscox}[1]{\setlength{\Laped}{2pt-#1pt}\ce{La_{\strip@pt\Laped}Sr_{#1}CuO$_4$}}
\newcommand{\lscoxx}{La$_{\rm{2-x}}$Sr$_{\rm{x}}$CuO$_4$}

\makeatother
\newcommand{\ybco}{YBa$_2$Cu$_3$O$_{6+x}$}

\newcommand{\lsco}{La$_{2-x}$Sr$_x$CuO$_4$}
\begin{document}



\title{Crystal Symmetry of Stripe Ordered \lscox{0.12} }

\author{R. Frison}
\affiliation{Physik-Institut, Universit\"{a}t Z\"{u}rich, Winterthurerstrasse 190, CH-8057 Z\"{u}rich, Switzerland}

\author{J. K\"{u}spert}
\affiliation{Physik-Institut, Universit\"{a}t Z\"{u}rich, Winterthurerstrasse 190, CH-8057 Z\"{u}rich, Switzerland}

\author{Qisi~Wang}
\affiliation{Physik-Institut, Universit\"{a}t Z\"{u}rich, Winterthurerstrasse 190, CH-8057 Z\"{u}rich, Switzerland}

\author{O. Ivashko}
\affiliation{Deutsches Elektronen-Synchrotron DESY, Notkestra{\ss}e 85, 22607 Hamburg, Germany.}

\author{M.~v.~Zimmermann}
\affiliation{Deutsches Elektronen-Synchrotron DESY, Notkestra{\ss}e 85, 22607 Hamburg, Germany.}

\author{M. Meven}
\affiliation{RWTH Aachen University, Institut für Kristallographie, 52056 Aachen, Germany}
\affiliation{Jülich Centre for Neutron Science (JCNS) at Heinz Maier-Leibnitz Zentrum (MLZ), 85747 Garching, Germany}

\author{D.~Bucher}
\affiliation{Physik-Institut, Universit\"{a}t Z\"{u}rich, Winterthurerstrasse 
190, CH-8057 Z\"{u}rich, Switzerland}

\author{J. Larsen}
\affiliation{Department of Physics, Technical University of Denmark, DK-2800 Kongens Lyngby, Denmark}
\affiliation{Danish Fundamental Metrology A/S, Kogle All\'e 5, 2970 H{\o}rsholm, Denmark}

\author{Ch. Niedermayer}
\affiliation{Laboratory for Neutron Scattering and Imaging, Paul Scherrer Institut, CH-5232 Villigen PSI, Switzerland.}

\author{M.~Janoschek}
\affiliation{Physik-Institut, Universit\"{a}t Z\"{u}rich, Winterthurerstrasse 190, CH-8057 Z\"{u}rich, Switzerland}
\affiliation{Laboratory for Neutron and Muon Instrumentation, Paul Scherrer Institut, CH-5232 Villigen PSI, Switzerland}

\author{T.~Kurosawa}
\affiliation{Department of Physics, Hokkaido University - Sapporo 060-0810, 
Japan}
 
\author{N.~Momono}
\affiliation{Department of Physics, Hokkaido University - Sapporo 060-0810, 
Japan}
\affiliation{Department of Sciences and Informatics, Muroran Institute of Technology, Muroran 050-8585, Japan}

\author{M.~Oda}
\affiliation{Department of Physics, Hokkaido University - Sapporo 060-0810, 
Japan}

\author{N.~B.~Christensen}
\email{nbch@fysik.dtu.dk}
\affiliation{Department of Physics, Technical University of Denmark, DK-2800 Kongens Lyngby, Denmark}

\author{J.~Chang}

\affiliation{Physik-Institut, Universit\"{a}t Z\"{u}rich, Winterthurerstrasse 190, CH-8057 Z\"{u}rich, Switzerland}

\date{\today}

\begin{abstract}
We present a combined x-ray and neutron diffraction study of the stripe ordered superconductor \lscox{0.12}. The average crystal structure is consistent with the orthorhombic $Bmab$ space group as commonly reported in 
the literature. This structure however is not symmetry compatible with a second order phase transition into the stripe order phase, and, as we report here numerous Bragg peaks forbidden in the $Bmab$ space group are observed. We have studied and analysed these $Bmab$-forbidden Bragg reflections. Fitting of the diffraction intensities yields monoclinic lattice distortions that are symmetry consistent with charge stripe order. 
\end{abstract}

\pacs{}

\maketitle



\section{Introduction}{\label{sec_intro}}
The average crystal structure across the cuprate high-temperature superconducting phase diagrams
was determined early on by means of neutron and x-ray diffraction \cite{goodenough_sur-une-nouvelle_1973,cava_crystal_1987,garbauskas_single-crystal_1987,yan_structure_1987,muller-buschbaum_zur-kenntnis_1989,Radaelli_Structural_1994}. Although superconductivity 
in the cuprates is unlikely driven by phonons, the atomic lattice coordination still has  relevance.  For example, charge density waves (CDW) competing with superconductivity are associated with 
lattice strain waves distorting the lattice away from the average structure. In underdoped {\ybco} (YBCO), for example, the average structure is described by the space group $Pmmm$ whereas the charge ordering strain waves breaks the mirror symmetry of the CuO$_2$ bilayers  generating a supercell with {{the same space group}} $Pmmm$ symmetry \cite{Forgan_microscopic_2015}.\\
For La-based cuprates, however, the strain wave induced subgroup crystal structure remains unsolved. The discovery 
of thermal Hall effect in {\lscoxx} \cite{GrissonnancheNature2019,BoulangerNatComm2020,GrissonnancheNatPhys2020} has been interpreted in terms of chiral phonon excitations that would require specific crystal structures. 
While it seems established that the average structure of {\lscoxx} can be well described by the orthorhombic space group $Bmab$ (space group 64) \cite{Radaelli_Structural_1994,braden_characterization_1992},
increasing evidence suggests the presence of additional subtle structural distortions both in doped and un-doped {\lscoxx}. 
Forbidden Bragg reflections  (systematic extinctions) \cite{international-union-of-crystallography_space-group_2005} in the space group 64 have already been reported and in some cases interpreted as a consequence of a different --local-- crystal structure at the twin boundaries \cite{horibe_microstructure_1997,horibe_direct_2000,christensen_bulk_2014,jacobsen_neutron_2015}. 
Neutron diffraction experiments performed at room temperature on detwinned {\lco} (LCO) and very under-doped {\lscoxx} (LSCO) single crystals \cite{reehuis_crystal_2006} revealed the observation of weak symmetry forbidden Bragg reflections. The existence of such peaks was interpreted as a deviation from the orthorhombic symmetry $Bmab$ to a monoclinic $B2/m$, thus preserving lattice centering. Such results were later confirmed in similar experiments on lightly doped and twinned \lscox{0.05} reporting a weak but persistent monoclinic distortion reaching its maximum below 50 K and gradually decreasing, without vanishing {{through a first order phase transition}}, up to 250 K \cite{singh_evidence_2016}. More recently a reinvestigation of the LCO crystal symmetry \cite{sapkota_reinvestigation_2021} showed, along with the $B2/m$ peaks, evidence of the loss of lattice centering due to the observation of Bragg peaks with odd-odd indices in the ($hk0$) plane and weak signatures of the $B2/m$ monoclinic peaks up to 500~K. It has thus been  proposed \cite{sapkota_reinvestigation_2021} that there is a possible direct transition from the high temperature tetragonal phase to monoclinic structure.\\
\begin{figure*}
    \center{\includegraphics[width=0.999\textwidth]{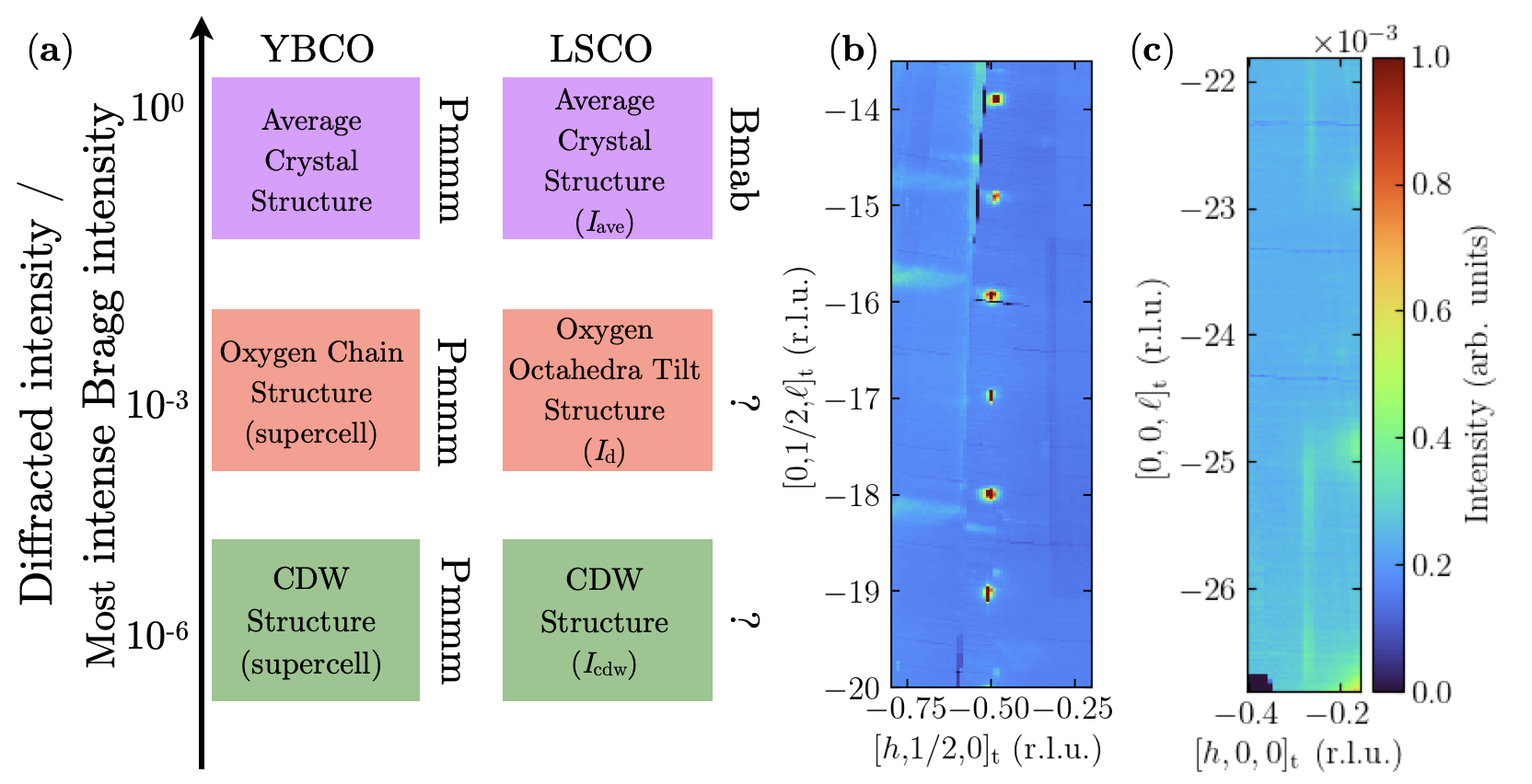}} 
    \caption{(a) Hierarchy of Bragg reflection intensity and crystal structure in YBCO and LSCO. Scattering intensity normalized to the most intense Bragg reflection is shown schematically. Intense fundamental lattice Bragg reflection are used for crystal structure determination. In both LSCO and YBCO, the charge density wave reflections are $10^{-6}-10^{-7}$ times weaker than the fundamental Bragg reflections. Oxygen chain order in YBCO and monoclinic distortion in LSCO manifest by moderately weak reflections {{in the ratio $10^{-2}-10^{-3}$ to that of fundamental Bragg reflections}}. For YBCO the crystal structure (including oxygen chain order) is determined to be $Pmmm$ and the charge density wave order generates a supercell with $Pmmm$ symmetry. The crystal structure of LSCO is not determined with the same precision. The average crystal structure defined by the strongest fundamental Bragg reflection is $Bmab$ (orthorhombic space group 64). However, the monoclinic and charge density wave reflections are inconsistent with  this average structure. The crystal symmetry of LSCO is therefore unsolved. {{(b) Portion of the reconstructed ($h,1/2,\ell$)$_{\rm{t}}$ plane showing some of $Bmab$-forbidden peaks. Gaussian fits of the $Bmab$-forbidden peaks along the $h$, $k$, and $\ell$ principal axes indicate that the correlation length $\xi$ along $a$ and $b$ directions is at least 50 unit cells, while along $c$ $\xi\gtrapprox10c$. Peaks of the kind ($o,0,e$) belong to the second twin component}}. (c) Section of the reconstructed ($h,0,\ell$)$_{\rm{t}}$ of reciprocal space along with CDW signal.}
    \label{figure1}
\end{figure*}
In parallel, CDW order in {\lscoxx} has been reported with wave vector ${\bm{q}}=(\simeq\pm1/4,0,1/2)$ \cite{Croft_Charge_2014,christensen_bulk_2014}. The emergence of CDW order can be interpreted as the consequence of a displacive continuous phase transition where the space group symmetries, before and after the transition, are  connected by a group-subgroup relation. Group theory \cite{toledano_the-landau_1987,toledano_extensions_2012,hatch_complete_2001} indicates which of the possible modulated displacement patterns are consistent with the observed CDW ordering wave vectors. Symmetry analysis  indicates that the stripe order observed in the {\lscoxx} system is not consistent with space group 64 as in this space group the $[1,0,0]_{\rm{t}}$ and $[0,1,0]_{\rm{t}}$ directions and all the CuO$_6$ octahedra are equivalent\footnote{The $[1,0,0]_{\rm{t}}$ and $[0,1,0]_{\rm{t}}$ directions are given by $[1,1,0]_{\rm{ort}}$ and $[1,\bar{1},0]_{\rm{ort}}$ in orthorhombic notation of space group $Bmab$.}. In La$_{1.85}$Ba$_{0.15}$CuO$_4$, for example, a direct tetragonal to monoclinic transition rather than tetragonal to orthorhombic \cite{moss_high-resolution_1987,Kajitani_Displacement_1987} has been proposed.  Experimental evidence \cite{reehuis_crystal_2006,jacobsen_neutron_2015,sapkota_reinvestigation_2021} shows that the {\lscoxx} system displays a hierarchy of lattice reflections as shown schematically in \fref{figure1}a. 
The strongest reflections defines the average structure ($I_{\rm{ave}}$). Weak $Bmab$-forbidden peaks with intensity  $I_{\rm{d}}\approx{\delta} I_{\rm{ave}}$ correspond to subtle lattice distortions {with $\delta$ ranging from $10^{-3}$ to $10^{-2}$} \fref{figure2}(a-f).
Finally, there are charge order induced strain wave reflections for which $I_{\rm{cdw}}\approx10^{-6}$-$10^{-7}I_{\rm{ave}}$ \fref{figure1}(b-c).  It is therefore important to solve the subgroup crystal structure problem accounting for the observed, coexisting, weak structural distortions.\\ Here we analyse the deviations from the average structure in a {\lscox{0.12}} crystal. We have carried out neutron and x-ray single-crystal diffraction (XRD) experiments. In the former the crystal was not detwinned, whereas in the latter, uniaxial pressure was applied along a copper-oxygen bond direction ($a_{\rm{t}}$) to minimize twinning effects.  We performed a systematic study of the symmetry forbidden Bragg peaks of the average structure. Our results are analysed and  discussed by identifying subgroups of the established average orthorhombic ($Bmab$) structure consistent consistent with the observed forbidden Bragg peaks, and via crystal structure refinement of the model candidates to identify the space group providing the best fit of the observed $Bmab$-forbidden Bragg peaks.

%
\section{Methods}{\label{sec_exp}}
We performed neutron diffraction experiment using a 5~mm $\times\oslash$5~mm {\lscox{0.12}} single crystal (T$_{\rm{c}}$ = 27~K) grown by the travelling solvent floating zone method \cite{nakano_correlation_1998,chang_tuning_2008}. Neutron diffraction data were collected at the  HEiDi Single crystal diffractometer at neutron source FRM-II of Heinz Maier - Leibnitz Zentrum (MLZ) in Garching near Munich using an Erbium filter with $\lambda=0.7094$~\AA~ and $q_{\rm{max}}=2\sin(\theta)/\lambda$ = 0.97 \AA$^{-1}$. 
For the x-ray experiment{s on the same crystal batch}, uniaxial pressure was applied \textit{ex-situ}, as described in Ref.~\onlinecite{choi_disentangling_2020}, along a Cu-O bond direction (${\bm{a}}_{\rm{t}}$ or ${\bm{b}}_{\rm{t}}$), to minimize orthorhombic twinning effects. X-ray diffraction data collection was performed at the P21.1 beamline at PETRA-III (Hamburg) synchrotron using $\lambda=0.122$~\AA~ in combination with a Perkin Elmer or a Dectris Pilatus 100K CdTe detector. 
Data indexing and integration was performed using XDS \cite{kabsch_it-xds_2010}.
Crystal structure refinement was done using Shelxl \cite{sheldrick_a-short_2008} and structure factor calculation of the distorted superstructure were performed using the {\textsc{fullprof suite}} \cite{rodriguez-carvajal_recent_1993}. {{Throughout the text, reciprocal space is indexed according to the high temperature tetragonal (HTT) structure as ($h,k,\ell)_{\rm{t}}$, or  according to the average low temperature orthorhombic structure as ($h,k,\ell)_{\rm{o}}$. The two indexing notations are connected by ($h,k,\ell)_{\rm{o}}=R_{\frac{\pi}{4}}$($h,k,\ell)_{\rm{t}}$ where $R_{\frac{\pi}{4}}$ is a matrix rotation around the $(0,0,1)$ axis. The choice of adopting the two indexing schemes reflects the fact that, throughout the existing literature, charge stripe order in LSCO is indicated in tetragonal notation, while distortions away from space group 64 is best described in orthorhombic notation.}}

\begin{figure*}
    \center{\includegraphics[width=0.999\textwidth]{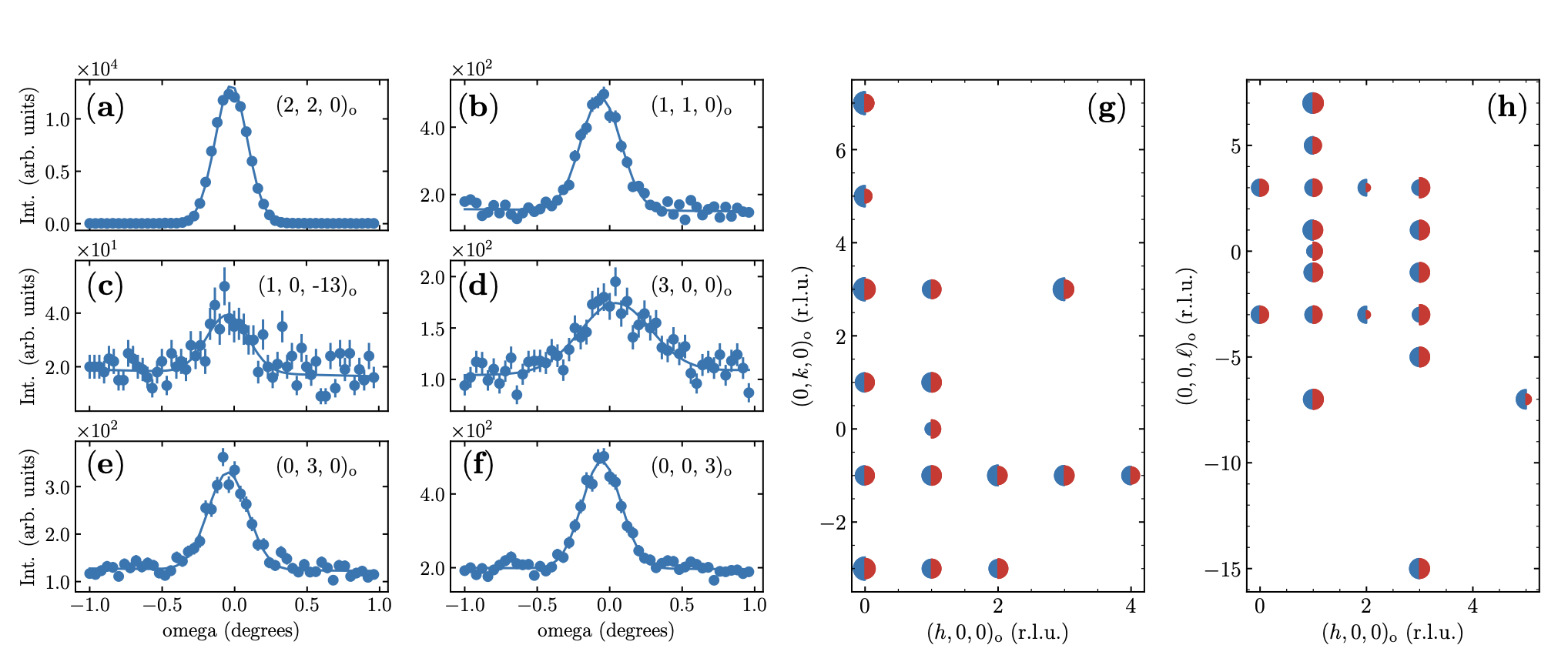}} 
    \caption{Allowed and forbidden Bragg peaks measured on \lscox{0.12} by neutron diffraction. (a-f)  Bragg peaks  indexed using orthorhombic $Bmab$ notation. (a) allowed Bragg reflection, (2,2,0)$_{\rm{o}}$; (b-f) Forbidden Bragg peaks of the kind ($o$,$o$,0)$_{\rm{o}}$, ($o$,0,$o$)$_{\rm{o}}$, ($o$,0,0)$_{\rm{o}}$, (0,$o$,0)$_{\rm{o}}$, and (0,0,$o$)$_{\rm{o}}$ with $o$ being an odd integer (see also text). The line through the data points is a gaussian fit to guide the eye. 
    (g,h) Observed (blue half-circle) and fitted (red half-circle) neutron diffraction intensities of $Bmab$-forbidden peaks using the $P2/m$ model  for the $(h,k,0)$ and $(h,0,\ell)$ planes. The radius of the semi-circles is proportional to the intensity of the corresponding Bragg peaks.}
    \label{figure2}
\end{figure*}

\section{Results}
\begin{figure*}
    \center{\includegraphics[width=0.999\textwidth]{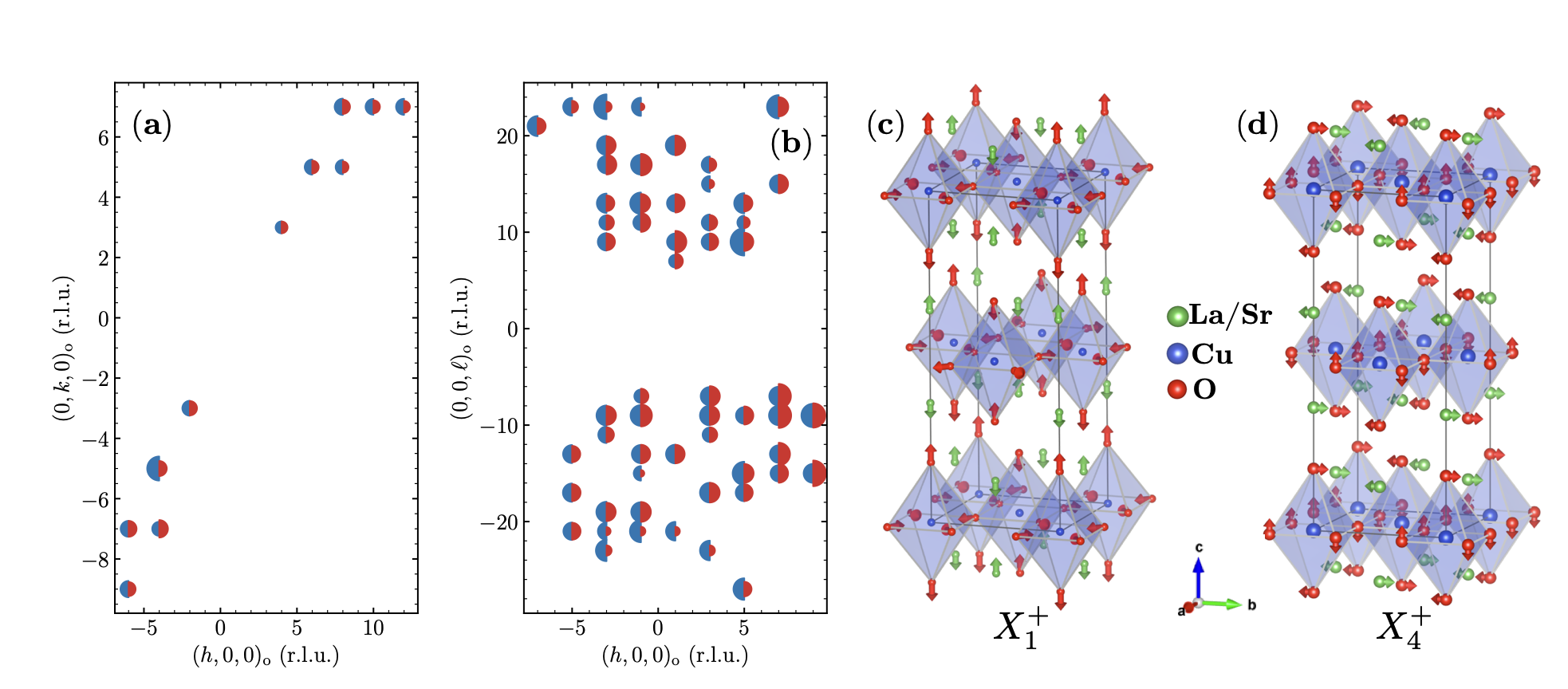}} 
    \caption{(a,b) Observed (blue half-circle) and fitted (red half-circle) x-ray diffraction intensities of $Bmab$-forbidden peaks using the $B2/m$ model  for the $(h,k,0)$ and $(h,0,\ell)$ planes. The radius of the semi-circles is proportional to the intensity of the corresponding Bragg peaks. (c,d) Representation of the atomic displacement motifs according to distortion modes (c) $X^+_1$ and (d) $X^+_4$ for an undistorted unit cell, the magnitude of the displacements have been exaggerated to make them visible.}
    \label{figure3}
\end{figure*}

{\label{sec_res}}
CDW stripe order manifests by reflections at $\bm{Q}=\bm{\tau}+(\delta,0,1/2)_{\rm{t}}$ with $\delta\simeq 1/4$ and $\bm{\tau}$ being a fundamental Bragg position. 
\fref{figure1}b,c display sections of the reconstructed reciprocal space probed by x-ray diffraction around $(1/4,0,\ell)_{\rm{t}}=(1/4,1/4,\ell)_{\rm{o}}$ and $(-1/2,1/2,\ell)_{\rm{t}}=(-1,0,\ell)_{\rm{o}}$ across multiple Brillouin
zones along the reciprocal $c$-axis. The out-of-plane charge order correlation length is small and hence the intensity, peaking at half-integer values of $\ell$, extends across the entire Brillouin zone.
The three-dimensional peaks at $(-0.5,0.5,o)_{\rm{t}}$, with $o$ being an odd integer, correspond to $(-1,0,o)_{\rm{o}}$ in $Bmab$ orthorhombic notation. In addition to the ($o,0,o$)$_{\rm{o}}$ reflections, weak Bragg peaks of the kind ($e,o,0$)$_{\rm{o}}$ 
with $e$ being an even integer are observed -- see \fref{figure3}(a-b). These reflection conditions cannot be explained even taking into account the presence of orthorhombic twin domains \cite{braden_characterization_1992}~{\footnote{Here orthorhombic twins implies an interchange of $h$ and $k$ indices.}}, and are therefore inconsistent with the space group $Bmab$. 
The observed reflection conditions are consistent with the monoclinic space group $B2/m(11)$, in agreement with previous results \cite{reehuis_crystal_2006,singh_evidence_2016}.\\ 
To exclude uniaxial pressure as the cause for symmetry reduction, we carried out a neutron diffraction experiment on a \lscox{0.12} crystal without uniaxial pressure applied. This dataset also displays weak reflections, with odd indices along the $h00$, $0k0$, and $00\ell$ axes, and of the kind ($e,e,o$)$_{\rm{o}}$, ($o,o,0$)$_{\rm{o}}$, ($e,o,0$)$_{\rm{o}}$, and ($o,0,o$)$_{\rm{o}}$ which are inconsistent with the space group $Bmab$ (\fref{figure2}(b-f)) and  cannot be explained by the twin law \cite{braden_characterization_1992}. In this case the observed reflection conditions indicate the space group $P2_1$ (4), in agreement with recent observations \cite{sapkota_reinvestigation_2021}. We note, however, that in this case the $(h,0,0)$ condition imposed by the space group $P2_1$, is masked by the presence of orthorhombic twins. Thus also the space group $P2/m$ is a plausible structure. {{Reflection conditions for the various space groups are reported in \tref{tab_sg}.}}
Before attempting a finer crystal structure refinement, we notice that  the x-ray and neutron diffraction experiments provide some overlap of ``forbidden'' Bragg peaks. Yet, the two datasets are not identical and hence are analysed separately. 

\section{Analysis}
{{For the neutron dataset, we performed refinements using the space groups $P2_1$ and $P2/m$ obtaining R=0.0974, and  R=0.0776 respectively, see \tref{lsco_n_lowsymm}.  For the x-ray dataset we tested the monoclinic space groups $Bm$ and $B2/m$, obtaining respectively R=0.081 and  R=0.077, see \tref{lsco_x_lowsymm}. In all these cases the intensity of the $Bmab$-forbidden Bragg peaks is underestimated. Further the Wilson statistic $\langle|E^2-1|\rangle$ is 1.3 and 1.5 for the neutron and x-ray case, indicating the presence of a centrosymmetric structure. Single crystal structure refinements favor the $Bmab$ space group for both our x-ray and neutron diffraction experiments.}} 
Therefore to provide a better fit to the forbidden peaks, we opted for partitioning the total intensity as $I_{\rm{tot}}\simeq I_{\rm{ave}}+I_{\rm{d}}$ where subscripts stand for total, average and distortion, respectively. Our working hypothesis is that the average structure is equivalent to the $Bmab$ space group and the weaker distortions represent small, static, correlated --symmetry breaking-- atomic displacements  away from the average structure.{\footnote{In the limit of small displacements we aim to produce negligible interference effects with the average structure. This is justified given the observed ratio between the $I_{\rm{ave}}$ $I_{\rm{d}}$ components.}}. The structural distortion component is further described in terms of modes superposition. Each mode is a collective correlated atomic displacements pattern fulfilling specific symmetry properties given by the irreducible representations (irreps) of the undistorted parent high-symmetry space group \cite{landau_statistical_1969,toledano_the-landau_1987,perez-mato_mode_2010}~\footnote{In this approach the irreps are fixed by symmetry, and the only free parameters are the amplitudes of the different modes, which can be refined in a standard least-squares fit while keeping fixed the other parameters. It is, thus, possible to determine which of the distortion modes contribute the most to the deviations from a parent average structure.
Under the assumption of the harmonic approximation, the distortion modes have a direct correspondence to the phonon eigenvectors.}.

To discuss the structural distortions in {\lscox{0.12}}, we start from the parent high-symmetry tetragonal $I4/mmm$ structure.
Orthorhombic structures manifest, in the first Brillouin zone,  at  $X$=($1/2$,$1/2$,0$)_{\rm{t}}$ \cite{miller_tables_1967,aleksandrov_successive_1987,Hatch_Phase_1989}. 
Group theory indicates that there are  seven displacement patterns (irreps)  consistent with this observed wave vector \cite{campbell_isodisplace_2006}: $X^+_1$, $X^+_2$, $X^+_3$, $X^+_4$, $X^-_2$, $X^-_3$, $X^-_4$.
The $Bmab$ structure, for example, corresponds to a CuO$_6$ octahedral tilt in the $[1,1,0]_{\rm{t}}$= $[0,1,0]_{\rm{o}}$ direction. This distortion pattern is described by the $X^+_3$ irreducible representation. In the same fashion, the monoclinic space groups, $P2/m$ and $B2/m$, are induced by the couplings $X^+_1\oplus X^+_3\oplus X^+_4$ and $X^+_1\oplus X^+_3$, respectively. The $X^+_1$ mode consists in a correlated displacement of the octahedral in-plane oxygens along the tetragonal in-plane axes and along the out-of-plane tetragonal axis of the octahedral apical oxygen atoms. The $X^+_4$ mode, instead, involves a tilt of the  CuO$_6$ octahedra around an in-plane axis, with octahedra in the first and second layer tilting out of phase.
The $X^+_1$ and $X^+_4$ distortion patterns are illustrated in \fref{figure3}c,d. We fitted the intensities of the $Bmab$-forbidden peaks optimising the mode amplitudes of the $X^+_1$ mode (x-ray) and $X^+_1$, $X^+_4$ modes (neutron), as these are the distinctive modes of the distorted structure.
As shown in \fref{figure2} and \fref{figure3}, reasonable agreement is obtained for both the neutron and x-ray diffraction experiments. The agreement factor $(\sum_i{|I^{\rm{obs}}_i-I^{\rm{calc}}_i|^2/\sigma_i})/(\sum_i{(I^{\rm{obs}}_i)^2/\sigma_i})$ for the two refinements is 7.4\% and 18.0\%, respectively. The mode amplitudes for each atomic site are given in \tref{lsco_ref_ampli}.\\ 
\begin{table}
\caption{Modulation amplitudes (in \AA) for each {\lscox{0.12}} HTT site. ``--'' marks amplitudes fixed to zero by symmetry; ``0'' marks amplitudes fixed manually to zero.}\label{lsco_ref_ampli}
\begin{ruledtabular}
\begin{tabular}{lcccc}
&\multicolumn{2}{c}{$P2/m$}&{$B2/m$}\\
\cline{2-4}
 Atom& $X^+_1$ & $X^+_4$& $X^+_1$ \\
 \hline
La &  0.0055(3)&  0&      -0.0010(9)\\
Sr &  0.0055(3)&  0&      -0.0010(9)\\
Cu &  --       & --&            --\\
O1 &  -0.057(1)& -0.2(1)&  -0.12(1)\\
O2 &  0.128(2) & 0.21(2)&   0.10(1)\\
\end{tabular}
\end{ruledtabular}
\end{table}

We now extend our symmetry analysis to include charge order.
Stripe order in LSCO is characterized by a uniaxial ordering vector $\bm{Q}\sim(1/4,0,1/2)_{\rm{t}}$ \cite{Croft_Charge_2014,christensen_bulk_2014,tranquada_evidence_1995}. 
This is contrast to YBCO, where a bi-directional charge density wave structure is reported \cite{ghiringhelli_long-range_2012,chang_direct_2012,achkar_distinct_2012}. The mono-directional stripe ordering vector of LSCO induces a further symmetry reduction which can be accounted for by a unit cell multiplication consistent with the ordering vector $\bm{Q}\sim(1/4,0,1/2)_{\rm{t}}$. As shown above, the existence of \textit{Bmab} forbidden Bragg peaks indicate monoclinic distortions which are described by specific irreps ($X^+_1$ and $X^+_4$). Group theory indicates \cite{campbell_isodisplace_2006} that the CDW wave vector corresponds to the irreps $B_1,B_2$. By coupling $B$ with the other irreps (determined on the basis of the average and monoclinic distortion), stripe order remains consistent with both $B2/m$ and $P2/m$ space groups.

\section{Discussion}
Different monoclinic structures are observed under ambient and uniaxial pressure application 
{suggesting} that uniaxial pressure influences the correlation of the weak lattice distortions. 
On the modelling side we {find}  
relatively high {fit} agreement factors, particularly for the x-ray dataset. 
We note that {\lscoxx} is characterized by intrinsic chemical disorder. In fact, while the average structure refinement confirms the $Bmab$ (LTO) structure as the best fitting model {(see \tref{lsco_ref_average})}, we found residual electron density peaks around the La/Sr position, which is not resolved refining the La(Sr) site occupation factor. It is thus expected that also the weak structural  distortion, and its corresponding intensity distribution, can  be affected by the presence of some occupational disorder. As a consequence, also  the fitness of our distortion model, which is only sensitive to the periodic features of the structure but responsible for the forbidden reflections, would be affected. The fitting model reproduces most of the modulations of the observed intensities \fref{figure3}a,b. As represented in \fref{figure3}c,d, the model describes correlated in-plane and out-of-plane displacements of the octahedral oxygen atoms such that in corner sharing octahedra the displacement has opposite sign.\\
Monoclinic distortions have also been reported for the parent {\lsco} and lightly doped compound \cite{reehuis_crystal_2006,singh_evidence_2016,sapkota_reinvestigation_2021}, where also
a thermal Hall effect has been reported \cite{GrissonnancheNature2019,BoulangerNatComm2020,GrissonnancheNatPhys2020} and interpreted in terms of chiral phonon excitations that would require specific crystal structures. In this context the connection between the observed monoclinic distortions and thermal Hall effect, could be tested by uniaxial pressure that seems to tune the former.\\
The present situation here described for {\lscoxx} shows some analogy and some difference with the case of YBCO. In YBCO 
different reciprocal space superstructures with periodicity $1/m$ ($m$=2,3,4,5,8) along the ${\bm{a}}^*$ axis have been reported \cite{zimmermann_oxygen-ordering_2003,chang_direct_2012}. Each of these corresponds to a specific ordering pattern of the chain-oxygens \cite{zimmermann_oxygen-ordering_2003}, thus with periodicity $m{\bm{a}}$, usually called ortho-$m$ structures. In these cases the multiplication of the unit cell in the $ab$-plane preserves the $Pmmm$ symmetry. The bi-axial charge order with ordering vectors $\bm{q}=(1/3,0,1/2)$ and $\bm{q}=(0,1/3,1/2)$ is produced by strain waves that break the bi-layer mirror symmetry \cite{Forgan_microscopic_2015}.  
The CDW modulated structure has been solved and described \cite{Forgan_microscopic_2015} using a superstructure with $Pmmm$ symmetry. Similarly, in {\lscoxx} octahedral tilts modes (and their superposition) induce structural distortions leading to a unit cell multiplication with, however, reduced symmetry. The monoclinic distortion, observed over a wide temperature range \cite{singh_evidence_2016}, is displaying long-range correlations along all principal crystal axes. 
The charge stripe order that is, by contrast, extremely weakly correlated across the CuO$_2$ layers. The two sets of distortions (charge stripe order and monoclinic) are therefore not directly linked. Yet, future experiments should address whether the monoclinic distortion interacts with superconductivity. It should be addressed, for example, whether the competition between stripe order and superconductivity is channeled through mutual interaction with the monoclinic distortions. Overall, our structural analysis suggests that the weak monoclinic lattice distortions are a necessary condition for charge stripe order in \lsco.\\

\section{Conclusions}
In summary, we have carried out a neutron and x-ray diffraction study to resolve the crystal structure underpinning charge stripe order in the high temperature superconductor \lscox{0.12}. The average orthorhombic $Bmab$ structure is symmetry inconsistent with the unidirectional charge order. We therefore analysed atomic distortions away from the average structure that manifest by  weak $Bmab$-forbidden Bragg peaks. We infer monoclinic $P2/m$ in absence of uniaxial pressure and $B2/m$ when uniaxial pressure along the copper-oxygen bond is applied. The $B2/m$ monoclinic space group is also preserved after coupling with the stripe order CDW distortion mode. We therefore conclude that weak monoclinic lattice distortions are an necessary precondition for the emergence of stripe order in \lsco.

\appendix

\section{Reflection conditions for the various space groups}

\begin{table*}
\caption{Space groups notations and their reflection conditions. Conditions are abbreviated assuming the expression is an even number \cite{international-union-of-crystallography_space-group_2005}.}\label{tab_sg}
\begin{ruledtabular}
\begin{tabular}{cccccccccc}
\multicolumn{2}{c}{Space Group}&\multicolumn{8}{c}{Reflection conditions}\\
\cline{1-2}\cline{3-10}
Symbol& No.& $hk\ell$& $hk0$& $h0\ell$& $0k\ell$& $hh\ell$& $h00$& $0k0$& $00\ell$\\  
\hline
$I4/mmmm$ & 139& $h+k+\ell$& $h+k$& & $k+\ell$& $\ell$& & $k$& $\ell$\\
$Bmab$ & 64& $h+\ell$& $h,k$& $h,\ell$& $\ell$& & $h$& $k$& $\ell$\\
$B2/m(11)$ & 12& $h+\ell$& $h+\ell$& $h+\ell$& $\ell$& & $h$& & \\
$P2/m(11)$ & 10& & & & & & & & \\
$Bm(11)$ & 8& $h+\ell$& $h+\ell$& $h+\ell$& $\ell$& & $h$& & \\
$P2_1(11)$ & 4& & & & & & $h$& & \\
\end{tabular}
\end{ruledtabular}
\end{table*}

\section{Crystal structure average structure refinement}
Results of the average structure refinement are provided in \tref{lsco_ref_average}. Both the x-ray and neutron diffraction yield an average $Bmab$ crystal structure.\\
Results of the refinements aimed at including the $Bmab$-forbidden peaks, hence capturing the structural distortion, using the space groups $P2_1$, $P2/m$ for the neutron dataset are reported in \tref{lsco_n_lowsymm}, and using the space groups $Bm$ and $B2/m$ for the x-ray dataset are reported in \tref{lsco_x_lowsymm}.  
\begin{table*}
\caption{Top: Positional and thermal parameters of {\lscox{0.12}} as obtained from the structure refinement of the neutron (top) and x-ray (bottom)  diffraction datasets using the orthorhombic $Bmab$ setting.\label{lsco_ref_average}}
\begin{ruledtabular}
{\lscox{0.12}} at 2 K, neutron diffraction $\lambda$=0.794~\AA : $a$=5.34(4)~\AA~ $b$=5.37(7)~\AA~ $c$=13.22(0)~\AA ~$\alpha$=$\beta$=$\gamma$=90~deg; Extinction coefficient = 0.037(6), twin fraction = 0.204(4); R=5.80\%, wR2=17.05\%, GooF=1.1.\\
\begin{tabular}{lcccccccccccc}
 Atom& site& $x$& $y$& $z$& $U_{11}$& $U_{22}$& $U_{33}$& $U_{23}$& $U_{13}$& $U_{12}$& $U_{\rm{eq}}$& Occ.\\
 \hline
 La& $8f$&        0&  -0.00610(11)&   0.36074(5)&     0.0013(4)&   0.0009(3)&   0.0015(3)&   0.00007(13)&   0&   0&    0.00121(19)& 0.875\\
 Sr& $8f$&        0&  -0.00610(11)&   0.36074(5)&     0.0013(4)&   0.0009(3)&   0.0015(3)&   0.00007(13)&   0&   0&    0.00121(19)& 0.125\\
 Cu& $4a$&        0&   0&   0&     0.0031(6)&   0.0023(5)&   0.0023(4)&   0.00023(18)&   0&   0&    0.0026(2)& 1.00\\
 O1& $8e$&        1/4&   1/4&  -0.00583(7)&     0.0028(5)&   0.0030(4)&   0.0045(3)&   0&   0&  -0.0009(3)&    0.0034(2)& 1.00\\
 O2& $8f$&         0&   0.0282(3)&   0.18252(8)&     0.0083(5)&   0.0068(4)&   0.0022(4)&   0.0001(3)&   0&   0&    0.0057(2)& 1.00\\
\end{tabular}
{\lscox{0.12}} at 30 K, x-ray diffraction $\lambda$=0.122~\AA : $a$=5.31(9)~\AA~ $b$=5.33(9)~\AA~ $c$=13.17(9)~\AA ~$\alpha$=$\beta$=$\gamma$=90~deg;  Extinction coefficient = 0.48(5), twin fraction = 0.121(2); R=3.63\%, wR2=13.26\%, GooF=1.17.\\
\begin{tabular}{lcccccccccccc}
Atom& site& $x$& $y$& $z$& $U_{11}$& $U_{22}$& $U_{33}$& $U_{23}$& $U_{13}$& $U_{12}$& $U_{\rm{eq}}$& Occ.\\
 \hline
 La& $8f$&         0&  -0.00502(2)& 0.36094(2)&     0.00188(5)& 0.00243(6)& 0.00076(5)&  0.00005(1)&   0&   0&    0.00169(3)&  0.875\\
 Sr& $8f$&         0&  -0.00502(2)& 0.36094(2)&     0.00188(5)& 0.00243(6)& 0.00076(5)&  0.00005(1)&   0&   0&    0.00169(3)& 0.125\\
 Cu& $4a$&         0& 0& 0&     0.0029(2)& 0.0019(2)& 0.0023(1)& 0.00008(3)&   0&   0&    0.00234(5)& 1\\
 O1& $8e$&          1/4&    1/4&   -0.00491(7)&     0.0031(4)& 0.0027(4)& 0.0062(3)&   0&   0&   0.0003(2)&    0.0040(2)& 1\\
 O2& $8f$&          0&   0.0234(3)& 0.18330(13)&     0.0096(5)& 0.0088(4)& 0.0034(3)& -0.0016(3)&   0&   0&    0.0073(2)& 1\\
\end{tabular}
\end{ruledtabular}
\end{table*}

\begin{table*}
\caption{Positional and thermal parameters of {\lscox{0.12}} as obtained from the structure refinement of the neutron diffraction datasets using the $P2_1$ (top) and $P2/m$ (bottom).\label{lsco_n_lowsymm}}
\begin{ruledtabular}
{\lscox{0.12}} at 2 K, neutron diffraction $\lambda$=0.794~\AA : $a$=5.34(4)~\AA~ $b$=5.37(7)~\AA~ $c$=13.22(0)~\AA ~$\alpha$=$\beta$=$\gamma$=90~deg; $P2_1$ symmetry, Extinction coefficient = 0.037(6), twin fraction = 0.204(4); R=9.74\%,  wR2=25.17\%,  GooF= 1.535.\\
\begin{tabular}{lcccccc}
 Atom& site& $x$& $y$& $z$& $U_{\rm{eq}}$& Occ.\\
 \hline
La& $2a$&        0.7547(6)&  -0.0004(5)&  0.3576(2)&      0.00037& 0.875\\
Sr& $2a$&        0.7547(6)&  -0.0004(5)&  0.3576(2)&      0.00037& 0.125\\
La& $2a$&        0.2460(6)&  -0.0104(5)&  0.8637(2)&      0.00006& 0.875\\
Sr& $2a$&        0.2460(6)&  -0.0104(5)&  0.8637(2)&      0.00006& 0.125\\
La& $2a$&        0.7464(7)&  0.0112(5)&   0.6424(2)&      0.00054& 0.875\\
Sr& $2a$&        0.7464(7)&  0.0112(5)&   0.6424(2)&      0.00054& 0.125\\
La& $2a$&        0.2535(6)&  0.0017(5)&   0.1357(2)&      0.00053& 0.875\\
Sr& $2a$&        0.2535(6)&  0.0017(5)&   0.1357(2)&      0.00053& 0.125\\
Cu& $2a$&        0.7498(6)&  -0.0039(9)&  0.0001(2)&      0.001& 1.0\\
Cu& $2a$&        0.251(1)&   0.007(2)&    0.5003(5)&      0.0168& 1.0\\
O& $2a$&         0.002(1)&   0.751(4)&    0.0055(6)&      0.010(1)& 1.0\\
O& $2a$&         0.4988(8)&  0.7504(8)&   0.5033(4)&      0.0014(6)& 1.0\\
O& $2a$&         0.502(1)&   0.251(1)&    -0.0096(6)&     0.0086(9)& 1.0\\
O& $2a$&        -0.0002(8)&  0.2505(9)&   0.4943(5)&      0.0018(7)& 1.0\\
O& $2a$&         0.755(1)&   0.031(1)&    0.1831(5)&      0.0097(9)& 1.0\\
O& $2a$&         0.2492(9)&  0.0244(9)&   0.6819(4)&      0.0059(6)& 1.0\\
O& $2a$&         0.754(1)&  -0.031(1)&    0.8179(4)&      0.0072(7)& 1.0\\
O& $2a$&         0.249(1)&  -0.026(1)&    0.3171(4)&     0.0078(8)& 1.0\\
\end{tabular}
{\lscox{0.12}} at 2 K, neutron diffraction $\lambda$=0.794~\AA : $a$=5.34(4)~\AA~ $b$=5.37(7)~\AA~ $c$=13.22(0)~\AA ~$\alpha$=$\beta$=$\gamma$=90~deg; $P2/m$ symmetry, Extinction coefficient = 0.027(4), twin fraction = 0.204(4); R=7.76\%,  wR2=20.95\%,  GooF= 1.295.\\
\begin{tabular}{lcccccc}
 Atom& site& $x$& $y$& $z$& $U_{\rm{eq}}$& Occ.\\
 \hline
La& $2m$&        0&  0.0059(7)&  0.3621(3)&        0.0004(--)&  0.4685\\
Sr& $2m$&        0&  0.0059(7)&  0.3621(3)&        0.0004(--)&  0.03125\\
La& $2n$&      1/2&  0.0060(8)&  0.8592(3)&         0.0021(3)&  0.4685\\
Sr& $2n$&      1/2&  0.0060(8)&  0.8592(3)&         0.0021(3)&  0.03125\\
La& $2m$&      0&    0.5064(7)&  0.1410(3)&         0.0024(4)&  0.4685\\
Sr& $2m$&      0&    0.5064(7)&  0.1410(3)&         0.0024(4)&  0.0312\\
La& $2n$&      1/2&  0.5061(6)&  0.6376(2)&         0.0011(4)&  0.4685\\

Cu& $1a$&        0&     0&     0&      0.001(--)& 0.25250\\
Cu& $1e$&      1/2&     0&   1/2&      0.001(--)& 0.25250\\
Cu& $1f$&      1/2&   1/2&     0&      0.0060(3)& 0.25250\\
Cu& $1g$&        0&   1/2&   1/2&      0.0060(3)& 0.25250\\
O& $4o$& 0.2495(4)&   0.7498(3)&    0.49527(13)&     0.00030(--)& 1.0\\
O& $4o$& 0.7509(6)&  0.7500(5)&   -0.0074(2)&       0.0089(3)& 1.0\\
O& $2m$&         0&   -0.0307(9)&    0.1826(4)&    0.0082(8)& 0.5\\
O& $2n$&         1/2&  -0.0269(8)&   0.6826(4)&    0.0052(7)& 0.5\\
O& $2m$&         0&   0.4750(9)&    0.3179(4)&     0.0058(7)& 0.5\\
O& $2a$&         1/2&  0.4691(9)&   0.8173(4)&     0.0064(7)& 0.5\\
\end{tabular}
\end{ruledtabular}
\end{table*}

\begin{table*}
\caption{Positional and thermal parameters of {\lscox{0.12}} as obtained from the structure refinement of the x-ray diffraction datasets using the $Bm$ (top) and $B2/m$ (bottom).\label{lsco_x_lowsymm}}
\begin{ruledtabular}
{\lscox{0.12}} at 30 K, x-ray diffraction $\lambda$=0.122~\AA : $a$=5.31(9)~\AA~ $b$=5.33(9)~\AA~ $c$=13.17(9)~\AA ~$\alpha$=$\beta$=$\gamma$=90~deg;  $Bm$ symmetry, Extinction coefficient = 0.48(5), twin fraction = 0.121(2); R=8.07\%, wR2=16.54\%, GooF=1.54.\\
\begin{tabular}{lcccccc}
 Atom& site& $x$& $y$& $z$&  $U_{\rm{eq}}$& Occ.\\
 \hline
La& $2a$&        0&  0.00457(5)&  0.36098(2)&    0.00112(7)& 0.4375\\
Sr& $2a$&        0&  0.00457(5)&  0.36098(2)&    0.00112(7)& 0.0625\\
La& $2a$&        0&  -0.00425(5)&  0.63904(2)&   0.00105(7)& 0.4375\\
Sr& $2a$&        0&  -0.00425(5)&  0.63904(2)&   0.00105(7)& 0.0625\\
La& $2a$&        0&  0.49517(5)&  0.86096(2)&    0.00111(6)& 0.4375\\
Sr& $2a$&        0&  0.49517(5)&  0.86096(2)&    0.00111(6)& 0.0625\\
La& $2a$&        0&  0.50464(5)&  0.13903(2)&    0.00104(7)& 0.4375\\
Sr& $2a$&        0&  0.50464(5)&  0.13903(2)&    0.00104(7)& 0.0625\\
Cu& $2a$&        0&  -0.0035(4)& -0.00030(17)&   0.00151(8)& 0.5\\
Cu& $2a$&        0&   0.4975(5)&  0.49968(17)&   0.00159(8)& 0.5\\
O& $4b$  0.7502(5)&   0.2514(10)&   0.0049(2)&   0.0047(3)&  1.0\\
O& $4b$& 0.2500(5)&  0.7517(10)&   0.4958(2)&    0.0054(3)&  1.0\\
O& $2a$&        0&  -0.0217(11)&   0.1813(2)&    0.0095(5)&  0.5\\
O& $2a$&         0&   0.0291(13)&  0.8171(3)&    0.0063(5)&  0.5\\
O& $2a$&         0&  0.5298(7)&   0.68289(16)&   0.0021(2)&  0.5\\
O& $2a$&         0&  0.4920(15)&    0.3193(4)&   0.0172(10)& 0.5\\
\end{tabular}
{\lscox{0.12}} at 30 K, x-ray diffraction $\lambda$=0.122~\AA : $a$=5.31(9)~\AA~ $b$=5.33(9)~\AA~ $c$=13.17(9)~\AA ~$\alpha$=$\beta$=$\gamma$=90~deg;  $B2/m$ symmetry, Extinction coefficient = 0.48(5), twin fraction = 0.121(2); R=7.66\%, wR2=16.15\%, GooF=1.39.\\
\begin{tabular}{lcccccc}
 Atom& site& $x$& $y$& $z$&  $U_{\rm{eq}}$& Occ.\\
 \hline
La& $4i$&        0&  -0.00457(4)&  0.36096(2)&     0.00127(5)&  0.4685\\
Sr& $4i$&        0&  -0.00457(4)&  0.36096(2)&     0.00127(5)&  0.03125\\
La& $4i$&        0&   0.49576(4)&  0.13903(2)&     0.00129(5)&  0.4685\\
Sr& $4i$&        0&   0.49576(4)&  0.13903(2)&     0.00129(5)&  0.03125\\
Cu& $2a$&        0&            0&           0&     0.00184(7)&  0.25250\\
Cu& $2d$&      1/2&          1/2&           0&     0.00185(7)&  0.25250\\
O& $8j$& 0.24995(12)& 0.25002(16)& -0.00433(7)&    0.00426(17)& 1.0\\
O& $4i$&         0&  0.0215(6)&    0.1806(2)&      0.0071(3)&   0.5\\
O& $2i$&         0&  0.5202(6)&    0.3196(2)&      0.0078(4)&    0.5\\
\end{tabular}
\end{ruledtabular}
\end{table*}

\begin{acknowledgments}
R.F., J.K., Q.W., D.B., J.C. thanks the Swiss National Science Foundation for support.
R.F. thanks Hans-Beat B\"urgi and Stefano Canossa for useful discussions. N.B.C. thanks the Danish Agency for Science, Technology, and Innovation for funding the instrument center DanScatt 
and acknowledges support from the Q-MAT ESS Lighthouse initiative. Neutron data were collected at the instrument HEiDi jointly operated by RWTH and JCNS at the MLZ within the JARA cooperation. We acknowledge DESY (Hamburg, Germany), a member of the Helmholtz Association HGF, for the provision of experimental facilities. Parts of this research were carried out at beamline P21.1.
\end{acknowledgments}

\clearpage
%

\end{document}